\documentclass{article}
\usepackage{spconf,graphicx}
\usepackage{amsmath,amssymb,amsfonts,amsthm}
\usepackage{subfig}
\usepackage{multirow}

\usepackage{euscript}
\usepackage[latin1]{inputenc}
\usepackage[T1]{fontenc}

\usepackage[english]{babel}
\usepackage{caption}
\usepackage{color}
\usepackage{booktabs}
\usepackage{times}
\usepackage{arydshln}
\usepackage{footmisc}
\usepackage{float}

\graphicspath{{hres/}}

\title{MODELING PLATE AND SPRING REVERBERATION USING A DSP-INFORMED DEEP NEURAL NETWORK}

\name{Marco A. Mart\'{i}nez Ram\'{i}rez, Emmanouil Benetos, Joshua D. Reiss\thanks{The Titan Xp GPU used for this research was donated by the NVIDIA Corporation. EB is supported by a RAEng Research Fellowship (RF/128). The \textit{spring} reverb samples were recorded with the help of Giulio Moro.}}
\address{Centre for Digital Music, Queen Mary University of London, UK}

\begin{document}
\ninept

{\textbf{IEEE Copyright Notice}\\
\copyright2020 IEEE2020 IEEE. Published in the IEEE 2020 International Conference on Acoustics, Speech, and Signal Processing (ICASSP 2020), scheduled for 4-9 May, 2020, in Barcelona, Spain. Personal use of this material is permitted. However, permission to reprint/republish this material for advertising or promotional purposes or for creating new collective works for resale or redistribution to servers or lists, or to reuse any copyrighted component of this work in other works, must be obtained from the IEEE. Contact: Manager, Copyrights and Permissions / IEEE Service Center / 445 Hoes Lane / P.O. Box 1331 / Piscataway, NJ 08855-1331, USA. Telephone: + Intl. 908-562-3966.
 }

\maketitle

\begin{abstract}
\textit{Plate} and \textit{spring} reverberators are electromechanical systems first used and researched as means to substitute real room reverberation. Currently, they are often used in music production for aesthetic reasons due to their particular sonic characteristics. The modeling of these audio processors and their perceptual qualities is difficult since they use mechanical elements together with analog electronics resulting in an extremely complex response. Based on digital reverberators that use sparse FIR filters, we propose a signal processing-informed deep learning architecture for the modeling of artificial reverberators. We explore the capabilities of deep neural networks to learn such highly nonlinear electromechanical responses and we perform modeling of \textit{plate} and \textit{spring} reverberators. In order to measure the performance of the model, we conduct a perceptual evaluation experiment and we also analyze how the given task is accomplished and what the model is actually learning.
\end{abstract}


\vspace{-2ex}

\section{Introduction}

\label{sec:intro}

Reverberation occurs when delayed and attenuated copies of the direct sound appear as reflections. Each reflection is frequency dependent and defined by the directivity of the sound source and the physical attributes of the reflecting surfaces \cite{zolzer2011dafx}. In the music and film industry, artificial reverberation was initially researched as a way of approximating the reflections occurring in room acoustics. This led to techniques that simulate reverberation, such as chamber, \textit{plate}, \textit{spring} and digital reverberators \cite{valimaki2012fifty}.

\textit{Plate} reverberation is based on a large metal plate which vibrates due to a moving-coil transducer attached to its centre. This transducer is fed with an amplified dry input signal and the plate vibrations are read by a pickup sensor and further amplified \cite{kuhl1958acoustical}. The \textit{plate} reverb sound is different from room acoustic reverberation and is characterized by a smooth noise-like response \cite{bilbao2009numerical}. \textit{Spring} reverberation is based on one or various helical springs suspended under low tension, attached to a magnetic bead and driven via an electromagnetic coupling \cite{parker2009spring}. The input audio source is transduced to spring vibrations which are read through a pickup sensor at the opposite end. The distinct sound of \textit{spring} reverb is due to the various types of vibrations that occur, transverse and longitudinal, which cause a peculiar combination of wave and dispersive propagation \cite{zolzer2011dafx}.

Although originally developed as substitutes for room reverberators, digital implementations of these devices have been widely researched due to their distinctive sound which has been of great interest among musicians, music producers and sound engineers. \textit{Plate} reverberation has been emulated with different approaches such as numerical simulation techniques, where a finite difference scheme \cite{bilbao2006physical, bilbao2007digital, arcas2010quality} or a modal description \cite{willemsen2017virtual} is derived from the differential equations that describe the motion of the plate; and hybrid digital filter-based algorithms \cite{abel2009emulation, greenblatt2010hybrid, lee2010approximating}, where convolutional impulse responses and feedback delay networks are used to model the desired impulse response. Similarly, modeling of \textit{spring} reverberation has been explored as wave digital filters \cite{abel2006spring}, to explicitly model the wave and dispersive propagation; numerical simulation techniques such as finite difference schemes \cite{bilbao2009virtual, parker2009spring, bilbao2013numerical}, and nonphysical modeling techniques \cite{valimaki2010parametric, parker2011efficient}, where chains of allpass filters and varying delay lines are used to approximate the dispersive and reverberant features of \textit{spring} reverb.

The modeling of these audio processors and their salient perceptual qualities remains an active research field. Their mechanical elements together with their analog circuitry yield a nonlinear and time-varying spatial system which is difficult to fully emulate digitally. Most of the methods are based on complete physical models or perceptual simplifications such as linearity and time-invariant behavior, thus, such models are not easily transferable to different artificial reverberators or cannot capture the full response of the system.

Deep learning architectures for black-box modeling of audio effects have been researched lately for linear effects such as equalization \cite{martinez2018end}; nonlinear memoryless effects such as tube amplifiers \cite{martinez2019modeling, damskagg2019deep, wright2019real}; nonlinear effects with temporal dependencies such as compressors \cite{hawley2019signaltrain}; and linear and nonlinear time-varying effects such as flanging or ring modulation \cite{martinez2019general}. Deep learning for dereverberation has become a heavily researched field \cite{feng2014speech, han2015learning}, although applying artificial reverberation or modeling \textit{plate} and \textit{spring} reverb with deep neural networks (DNN) has not been explored yet.

\begin{figure*}[t]
\hspace*{-0.75cm} 
\centering
\makebox[0pt]{\includegraphics[width=1.1\linewidth]{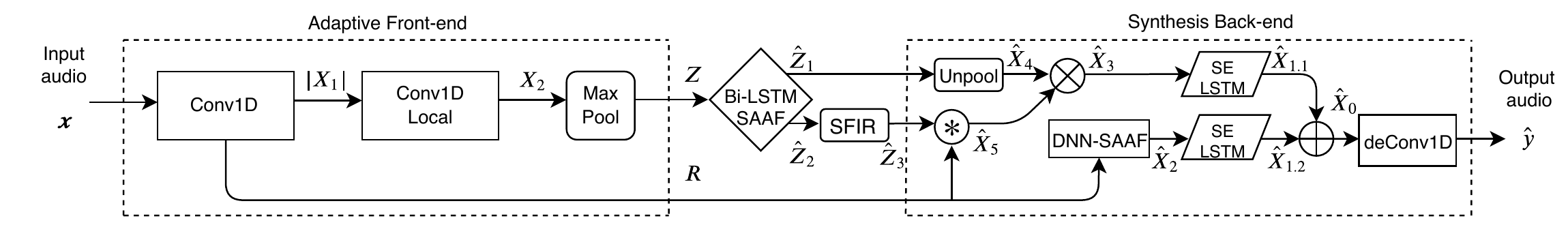}}
\caption{\label{fig:model}{Block diagram of the proposed model; adaptive front-end, latent-space and synthesis back-end.}}
\end{figure*}

Sparse FIR filtering has proven to be an efficient digital reverberation method \cite{rubak1999artificial, valimaki2012fifty}. Dispersive reflections are approximated via FIR filters with sparsely placed coefficients, which are often determined by a pseudo-random number sequence such as velvet noise \cite{jarvelainen2007reverberation}. We incorporate these methods to model noise-like and dispersive responses, such as those present in \textit{plate} and \textit{spring} devices.
\vbox{
Prior to this work, end-to-end DNNs have not yet been implemented to model artificial reverberators, i.e. learning from input-output data and applying the reverberant effect directly to the dry input audio. In this paper, we use convolutional, recurrent and dense layers together with sparse FIR (SFIR) filters and time-varying mixing gains which coefficients are learnt by the network. We explore whether a deep learning architecture is able to emulate \textit{plate} and \textit{spring} reverberators and we measure the performance of the model through a listening test. Both perceptual and objective evaluations indicate that the proposed model successfully simulates the electromechanical devices and performs better than other DNNs for modeling audio effects.
 \nopagebreak
\vspace{-2ex}
\section{Methods}\label{sec:methods}
}

\vspace{-2.5ex}
\subsection{Model}
\vspace{-1ex}

The model is completely based on time-domain input and works with raw and processed audio as input and output respectively. It is divided into three parts: adaptive front-end, latent-space and synthesis back-end. A block diagram is depicted in Fig. \ref{fig:model} and a more detailed structure can be seen online\footnote{\label{website}https://mchijmma.github.io/modeling-plate-spring-reverb/} together with the source code.

The \textbf{adaptive front-end} uses a filter bank architecture which learns the latent representation $\boldsymbol{Z}$ by performing time-domain convolutions with the input audio. It follows the same architecture as \cite{martinez2019general}, where it contains two convolutional layers, one pooling layer and one residual connection. The front-end is considered adaptive since its convolutional layers learn a filter bank for each modeling task and directly from the audio. The model learns long-term memory dependencies by having an input $\boldsymbol{x}$ which consists of the current audio frame $x$ concatenated with the $\pm 4$ previous and subsequent frames. These frames are of size $4096$ ($256$ ms) and sampled with a hop size of $50\%$.

The \textit{Conv1D} layer has $32$ one-dimensional filters of size $64$ and is followed by the \textit{absolute value} as nonlinear activation function. \textit{Conv1D-Local} has $32$ filters of size $128$, each filter is locally connected and uses the \textit{softplus} function as nonlinearity. This means we follow a filter bank architecture since each filter is only applied to its corresponding row in $|\boldsymbol{X}_{1}|$: the output of \textit{Conv1D}. The residual connection $\boldsymbol{R}$ is the corresponding row in $\boldsymbol{X}_{1}$ and is the frequency band decomposition of the current input frame $x$. This is due the output of each filter of \textit{Conv1D} can be seen as a frequency band. The \textit{max-pooling} layer is a moving window of size $64$, where the maximum value within each window corresponds to the output. All convolutional and pooling layers are time distributed, i.e. the same layer is applied to each of the $9$ input frames from $\boldsymbol{x}$. 

The \textbf{latent-space} has as its main objective to process $\boldsymbol{Z}$ into two latent representations, $\boldsymbol{\hat{Z}_1}$ and $\boldsymbol{\hat{Z}_2}$. The former corresponds to a set of envelope signals and the later is used to create the set of sparse FIR filters $\boldsymbol{\hat{Z}_3}$. It consists of two shared Bidirectional Long Short-Term Memory (Bi-LSTM) layers of $64$, $32$ units with the \textit{hyperbolic tangent} as activation function and its output is fed to two independent Bi-LSTM layers of $16$ units. Each of these layers is followed by a Smooth Adaptive Activation Function (SAAF) as the nonlinearity \cite{hou2017convnets}, obtaining in this way $\boldsymbol{\hat{Z}_1}$ and $\boldsymbol{\hat{Z}_2}$. SAAFs consist of piecewise second order polynomials which can approximate any continuous function and are regularized under a Lipschitz constant to ensure smoothness. As shown in \cite{martinez2019modeling}, SAAFs can be used as nonlinearities or waveshapers in audio processing tasks. 

We propose a \textit{SFIR} layer where we follow the constraints of sparse pseudo-random reverberation algorithms \cite{valimaki2012fifty}. Nevertheless, instead of using discrete coefficient values such as $-1$ and $+1$, each coefficient can take any continuous value within that range. Each one of the coefficients is placed at a specific index position within each interval of $T_s$ samples while all the other samples are zero. Thus, the \textit{SFIR} layer processes $\boldsymbol{\hat{Z}_2}$ by two independent fully connected (FC) layers of $1024$ units each. The FC layers are followed by a \textit{hyperbolic tangent} and \textit{sigmoid} function, whose outputs are the coefficient values and their index position respectively. To obtain the specific index position, the output of the \textit{sigmoid} function is multiplied by $T_s$ and a rounding down to the nearest integer is applied. This operation is not differentiable so we use an identity gradient as a backward pass approximation \cite{athalye2018obfuscated}. In order to have a high-quality reverberation, we use $2000$ coefficients per second \cite{rubak1999artificial}, thus, $T_s=8$ samples for a sampling rate of $16$ kHz.

The \textbf{synthesis back-end} uses the \textit{SFIR} output $\boldsymbol{\hat{Z}_3}$, the envelopes $\boldsymbol{\hat{Z}_1}$ and the residual connection $\boldsymbol{R}$ to synthesize the waveform and accomplish the reverberation task. It consists of an unpooling layer, a convolution and multiplication operation, a DNN with SAAFs (\textit{DNN-SAAF}), two Squeeze-and-Excitation \cite{hu2018squeeze} LSTM layers (\textit{SE-LSTM}) and a final convolutional layer.

Following the filter bank architecture: $\boldsymbol{\hat{X}}_{4}$ is obtained by upsampling $\boldsymbol{\hat{Z}}_{1}$ and the feature map  $\boldsymbol{\hat{X}}_{5}$ is accomplished by the locally connected convolution between the frequency band decomposition $\boldsymbol{R}$ and $\boldsymbol{\hat{Z}}_{3}$. The result of this convolution can be seen as explicitly modeling a frequency dependent reverberation response with the incoming audio. Furthermore, due to the temporal dependencies learnt by the Bi-LSTMs, $\boldsymbol{\hat{X}}_{5}$ is able to represent from the onset response the late reflections of the reverberation task. Then the feature map $\boldsymbol{\hat{X}}_{3}$ is the result of the element-wise multiplication of the reverberant response $\boldsymbol{\hat{X}}_{5}$ and the learnt envelopes $\boldsymbol{\hat{X}}_{4}$. The envelopes are applied in order to avoid audible artifacts between input frames \cite{jarvelainen2007reverberation}. 

Secondly, the feature map $\boldsymbol{\hat{X}}_{2}$ is obtained when the dynamic nonlinearites from the \textit{DNN-SAAF} block are applied to $\boldsymbol{R}$. The result of this operation consists of a learnt nonlinear transformation or waveshaping of the direct sound \cite{martinez2019modeling}. The \textit{DNN-SAAF} block consists of $4$ FC layers of $32$, $16$, $16$ and $32$ hidden units respectively. Each FC layer uses the \textit{hyperbolic tangent} as nonlinearity except for the last one, which uses a SAAF layer. 

Furthermore, we propose an \textit{SE-LSTM} block to act as a time-varying gain for $\boldsymbol{\hat{X}}_{2}$ and $\boldsymbol{\hat{X}}_{3}$. Since Squeeze-and-Excitation (SE) blocks explicitly and adaptively scale the channel-wise information of feature maps \cite{hu2018squeeze}, we incorporate an LSTM layer in the SE architecture in order to include long-term context from the input. Each \textit{SE-LSTM} builds on the architecture from \cite{kim2018sample}, it consists of an \textit{absolute value} operation and global average pooling operation followed by one LSTM and two FC layers of $32$, $512$ and $32$ hidden units respectively. The LSTM and first FC layer are followed by a rectifier linear unit, while the last FC layer uses a \textit{sigmoid} activation function. The \textit{absolute value} is incorporated before the global average pooling since the feature maps are based on time-domain waveforms. Each \textit{SE-LSTM} block process each feature map $\boldsymbol{\hat{X}}_{2}$ and $\boldsymbol{\hat{X}}_{3}$, thus, applying a frequency dependent time-varying mixing gain where outputs are added together in order to obtain $\boldsymbol{\hat{X}}_{0}$.

\begin{figure*}[t]

\begin{minipage}[b]{.25\textwidth}
  \centering
  \centerline{\includegraphics[width=1.\linewidth,height=0.12\textheight]{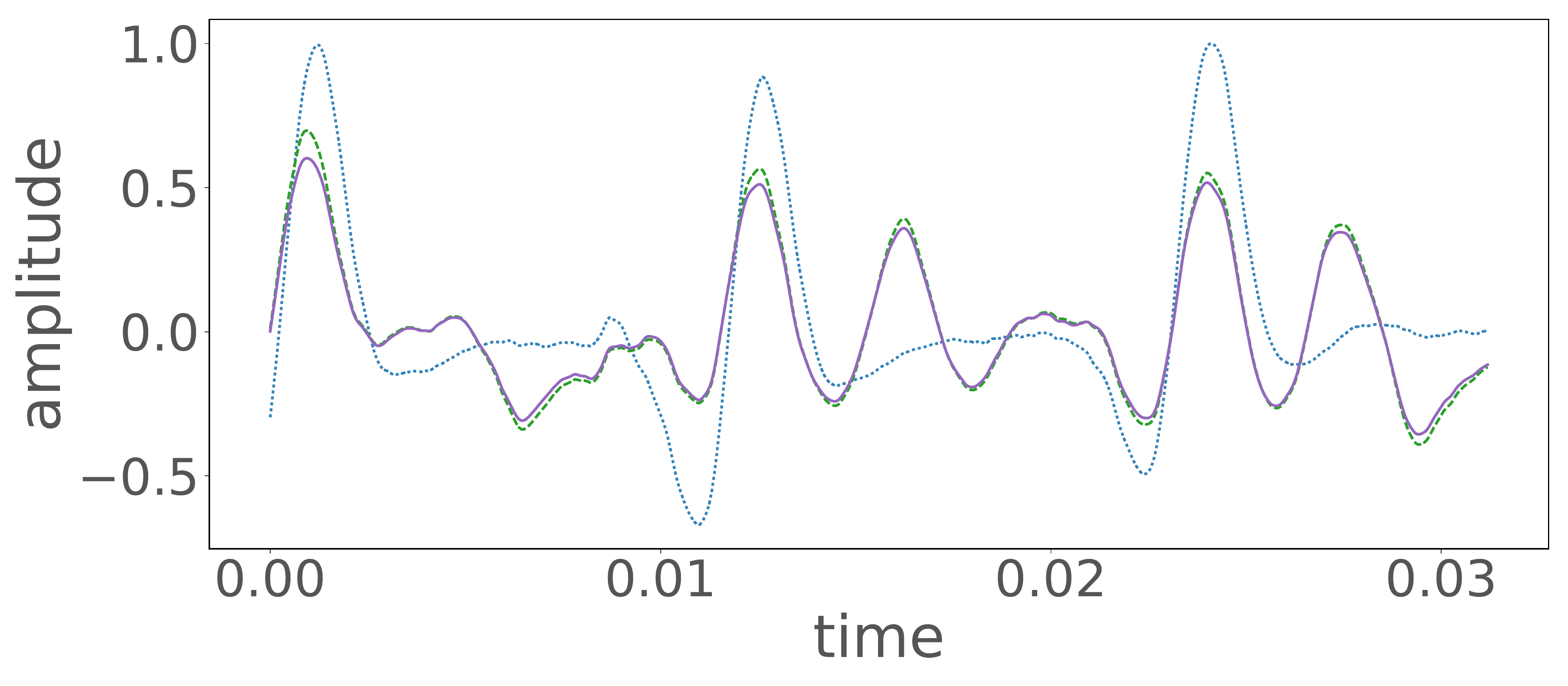}\label{fig:plate-frames}}
  \centerline{\includegraphics[width=1.\linewidth,height=0.12\textheight]{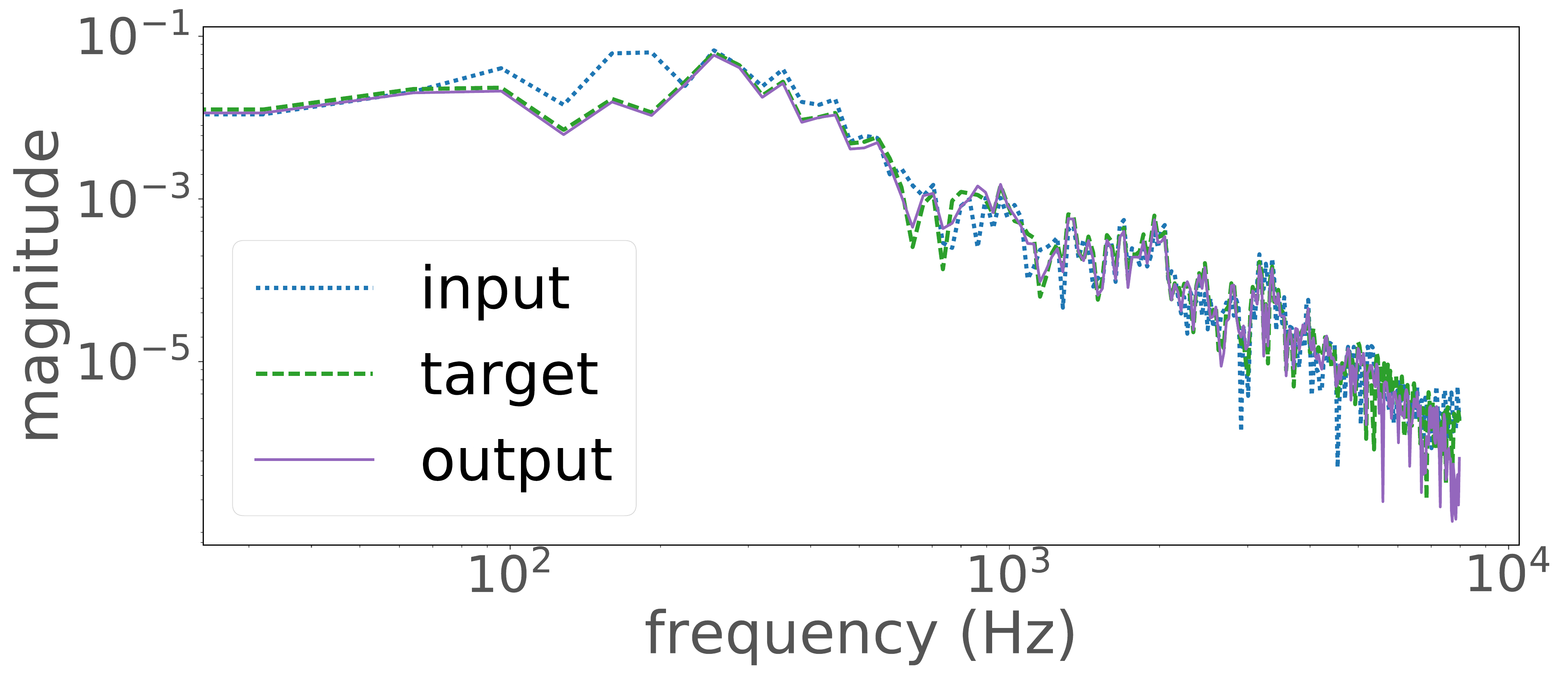}\label{fig:plate-fft}}
  \centerline{(a)}
\end{minipage}
\begin{minipage}[b]{0.24\textwidth}
  \centering
  \centerline{\includegraphics[width=1.\linewidth,height=0.2475\textheight]{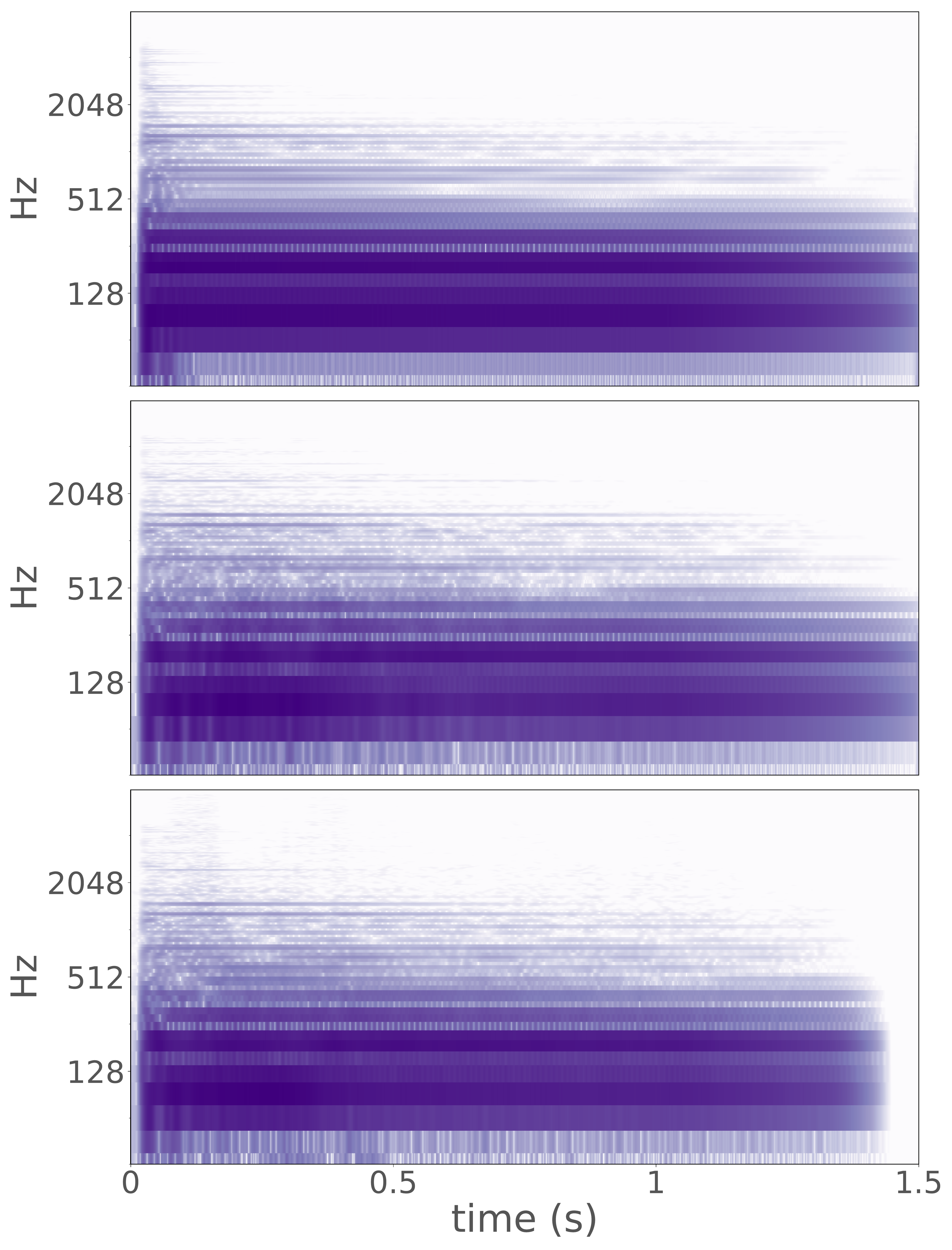}\label{fig:plate-spectogram}}
  \centerline{(b) }
\end{minipage}
\begin{minipage}[b]{.25\textwidth}
  \centering
  \centerline{\includegraphics[width=1.\linewidth,height=0.12\textheight]{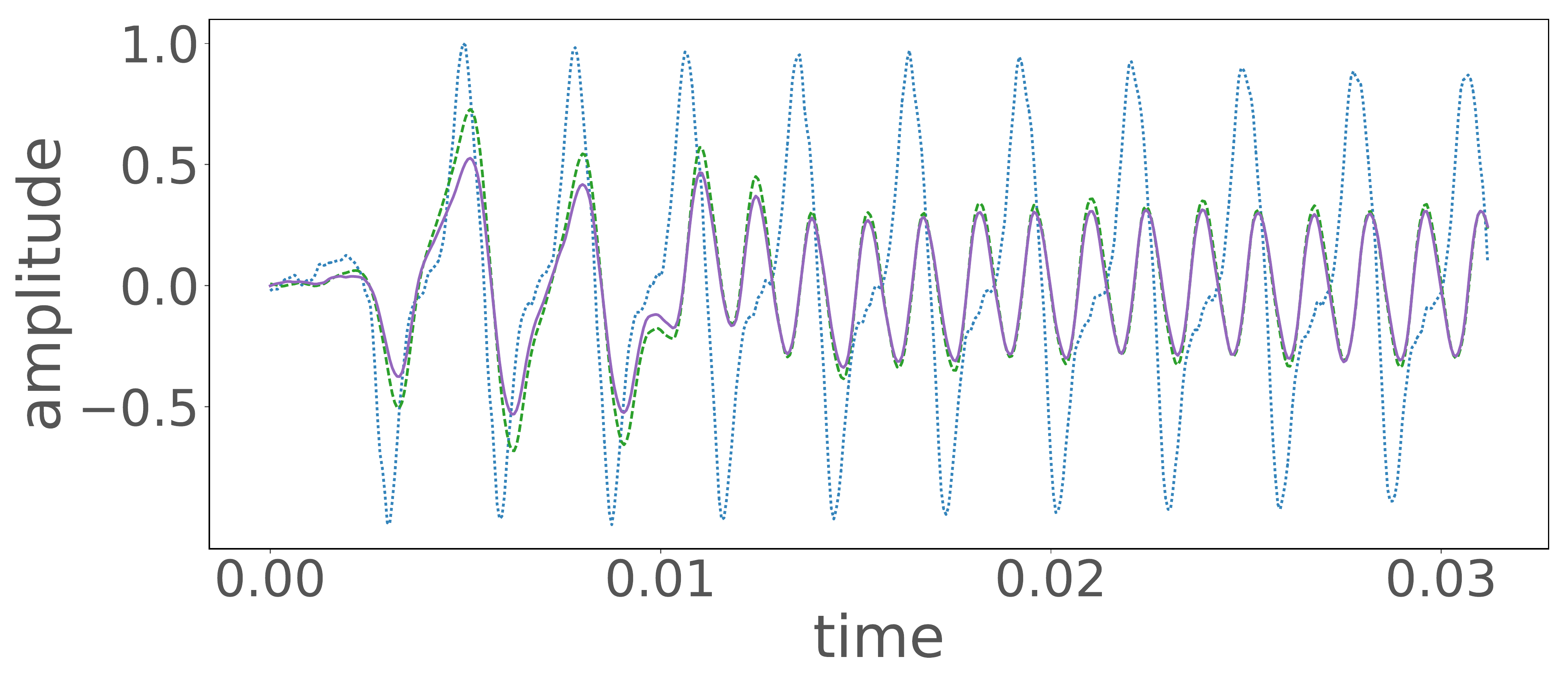}\label{fig:spring-frames}}
  \centerline{\includegraphics[width=1.\linewidth,height=0.12\textheight]{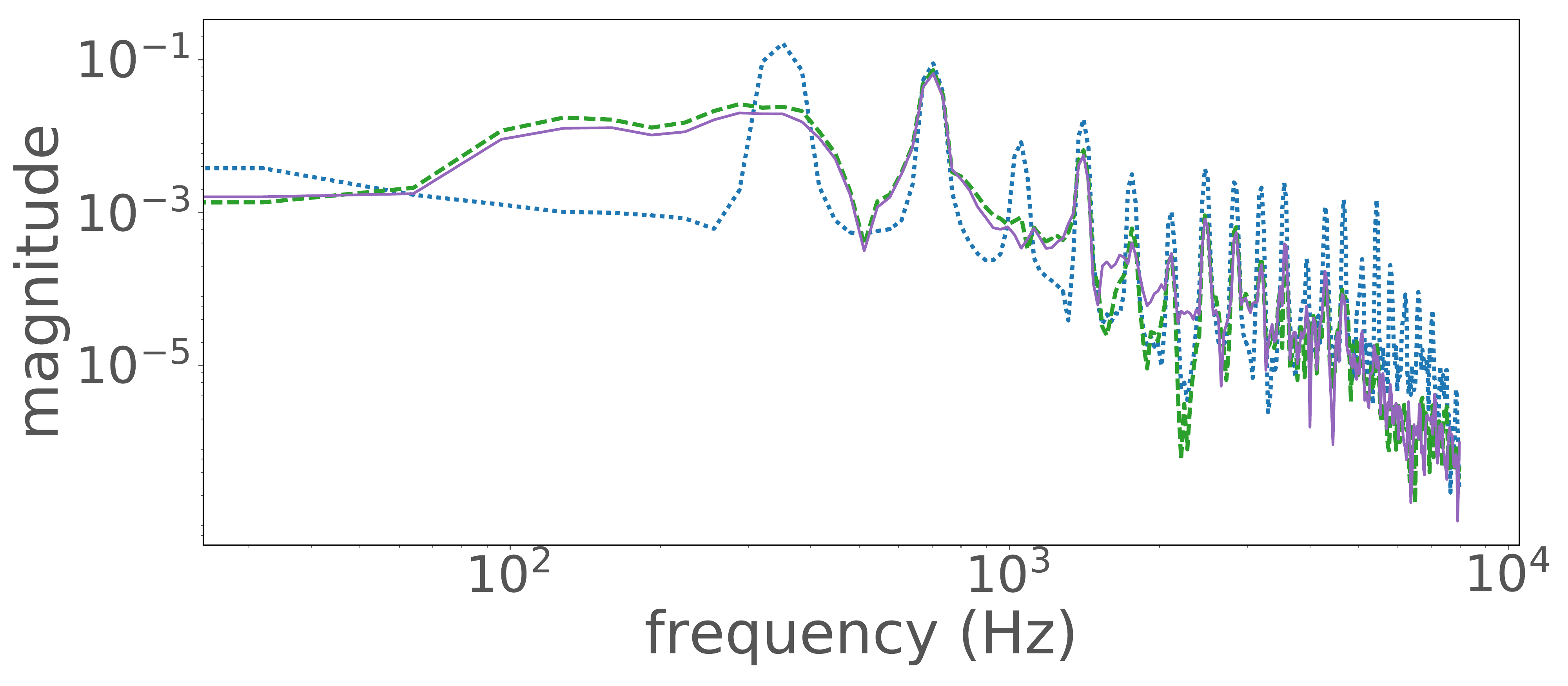}\label{fig:spring-fft}}
  \centerline{(c)}
\end{minipage}
\begin{minipage}[b]{0.24\textwidth}
  \centering
  \centerline{\includegraphics[width=1.\linewidth,height=0.2475\textheight]{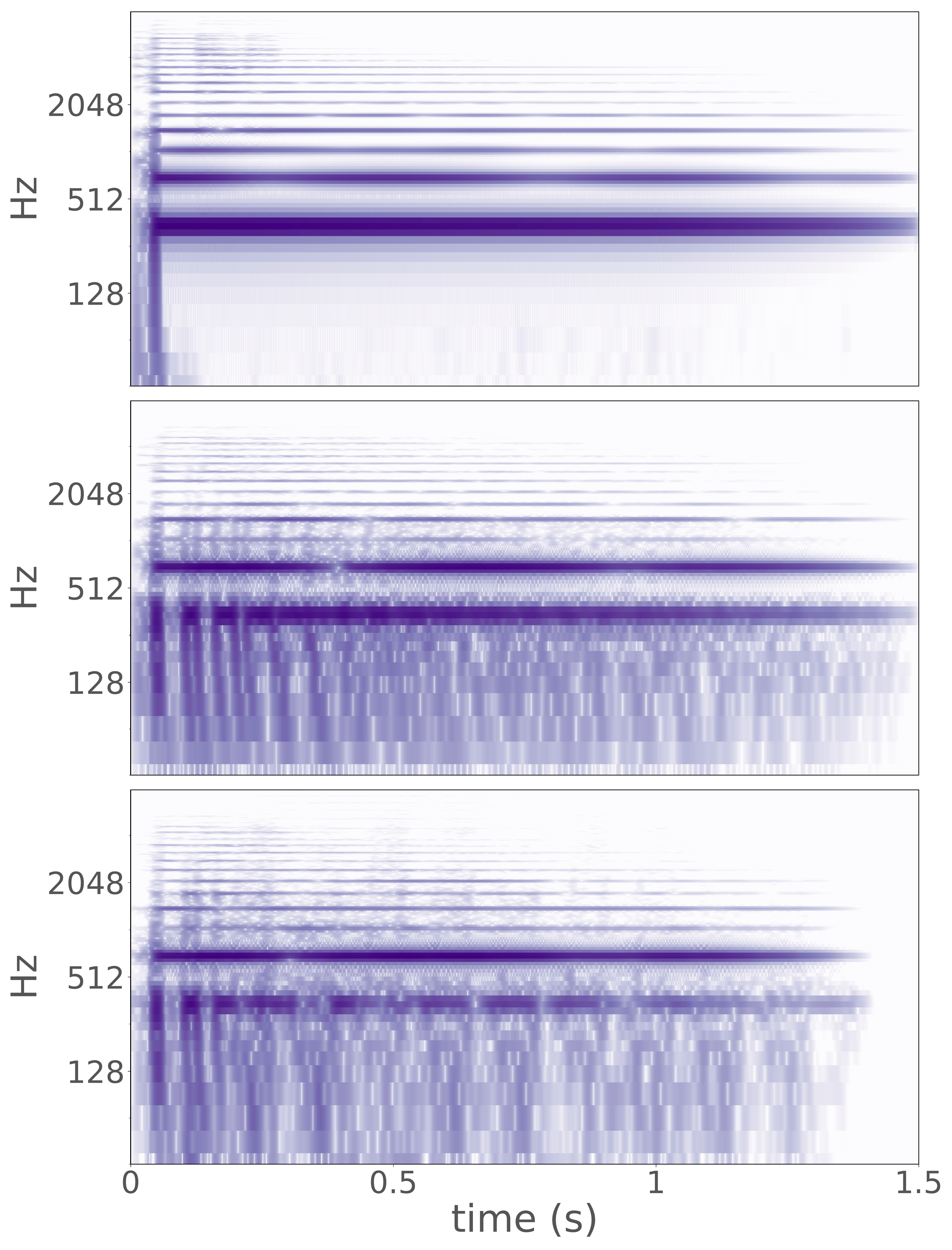}\label{fig:spring-spectogram}}
  \centerline{(d)}
\end{minipage}
\caption{Results for \textit{model-1} with the test dataset. \ref{fig:frames}a and \ref{fig:frames}c) show a segment of the input, target and output frames and their respective FFT magnitudes for \textit{plate} and \textit{spring} reverb, respectively. \ref{fig:frames}b) \textit{plate} and \ref{fig:frames}d) \textit{spring} reverb and from top to bottom: input, target and output spectrograms of the test samples; color intensity represents higher magnitude.}
\label{fig:frames}

\end{figure*}

The last layer corresponds to the \textit{deconvolution} operation which is not trainable since its filters are the transposed weights of \textit{Conv1D}. The complete waveform is synthesized using a \textit{hann} window and constant overlap-add gain. All convolutions are along the time dimension and all strides are of unit value. Overall, each SAAF is locally connected and each function consists of $25$ intervals between $-1$ to $1$ and each Bi-LSTM and LSTM have dropout and recurrent dropout rates of $0.1$.

\vspace{-2.5ex}

\subsection{Training}

\vspace{-1ex}

Prior to training the whole model, as an initialization step, only the weights of \textit{Conv1D} and \textit{Conv1D-Local} are trained. Thus, within an unsupervised learning task, the adaptive front-end is able to process and reconstruct both the dry audio $x$ and target audio $y$. This pretraining allows to have a better fitting when training for the reverberation task. During this step the unpooling layer of the back-end uses the time positions of the maximum values recorded by the \textit{max-pooling} operation. Once the front-end is initialized, all the weights of the convolutional, recurrent, dense and activation layers are trained following an end-to-end supervised learning task. The loss function to be minimized is based in time and frequency and described by:
\begin{equation}
loss = \alpha_1\cdot\textit{mae}(y,\hat{y}) + \alpha_2\cdot\textit{mse}(Y,\hat{Y})
\label{eq:1-loss}
\end{equation}

Where \textit{mae} is the mean absolute error and \textit{mse} is the mean squared error. $Y$ and $\hat{Y}$ are the log power magnitude spectra of the target and output respectively, and $y$ and $\hat{y}$ their respective waveforms. Prior to calculating the \textit{mae}, a pre-emphasis filter $H(z)=1-0,95z^{-1}$ is applied to $y$ and $\hat{y}$, in order to add more weight to high frequencies \cite{damskagg2019deep}. We use a 4096-point Fourier transform (FFT) to obtain $Y$ and $\hat{Y}$. In order to scale the time and frequency losses, we use $1.0$ and $1e-4$ as the loss weights $\alpha_1$ and $\alpha_2$ respectively. Explicit minimization in the frequency and time domains resulted crucial when modeling such complex responses.

For both training steps, \textit{Adam} \cite{kingma2014adam} is used as optimizer and we use an early stopping patience of $25$ epochs if there is no improvement in the validation loss. Afterwards, the model is fine-tuned further with the learning rate reduced by $25\%$ and also a patience of $25$ epochs. The initial learning rate is $1e-4$ and the batch size consists of the total number of frames per audio sample. We select the model with the lowest error for the validation subset.

\vspace{-2.5ex}

\subsection{Dataset}

\vspace{-1ex}

\textit{Plate} reverberation is obtained from the \textit{IDMT-SMT-Audio-Effects} dataset \cite{stein2010automatic}, which corresponds to individual 2-second notes and covers the common pitch range of various electric guitars and bass guitars. We use raw and \textit{plate} reverb notes from the bass guitar recordings. \textit{Spring} reverberation samples are obtained by processing the electric guitar raw audio samples with the \textit{spring} reverb tank \textit{Accutronics 4EB2C1B}.

For each reverb task we use $624$ raw and effected notes and both the test and validation samples correspond to $5\%$ of this subset each. The recordings are downsampled to $16$ kHz and amplitude normalization is applied. Also, since the \textit{plate} reverb samples have a fade-out applied in the last $0.5$ seconds of the recordings, we process the \textit{spring} reverb samples accordingly.

\begin{table}[b!]
\begin{center}
\caption{\textit{loss} values for \textit{plate} and \textit{spring} reverb models when tested with the test dataset.}

\begin{tabular}{cc||c|c|c}
\hline
 Reverb & model &\textit{mae} & \textit{mse} & \textit{loss} \\[1ex]
 \hline
\vspace{2pt}
plate
&\vspace{0pt} 1 &\vspace{0pt} \textbf{0.00214} &\vspace{0pt} \textbf{7.75815} &\vspace{0pt} \textbf{0.00292} \\
&\vspace{0pt} 2 &\vspace{0pt} 0.00316 &\vspace{0pt} 27.08704 &\vspace{0pt} 0.00587 \\
\hline
\textit{spring} 
&\vspace{0pt} 1 &\vspace{0pt} \textbf{0.00366} &\vspace{0pt} \textbf{9.43629} &\vspace{0pt} \textbf{0.00461}\\
&\vspace{0pt} 2 &\vspace{0pt} 0.00474 &\vspace{0pt} 33.09621 &\vspace{0pt} 0.00805 \\
\hline
\end{tabular}

\label{table:loss}
\end{center}
\end{table}


\vspace{-2ex}
\section{Results \& Analysis}

\label{sec:results}

\begin{figure}[t]
\begin{center}
\begin{minipage}{0.5\textwidth}
\begin{center}
\subfloat[]{\includegraphics[width=.49\linewidth, height=0.245\linewidth]{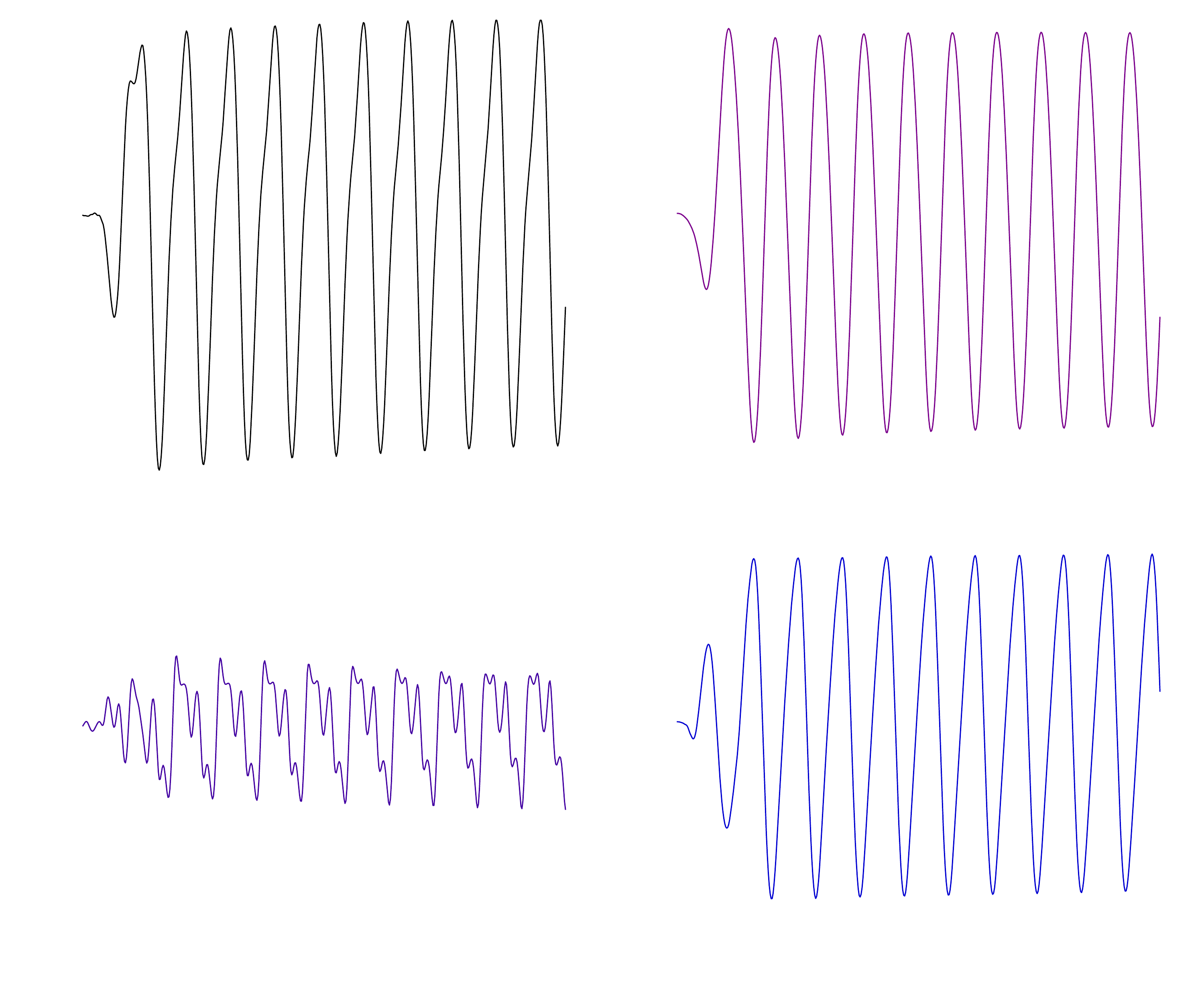}\label{fig:spring-x1}}\hspace{0ex}
\subfloat[]{\includegraphics[width=.49\linewidth, height=0.245\linewidth]{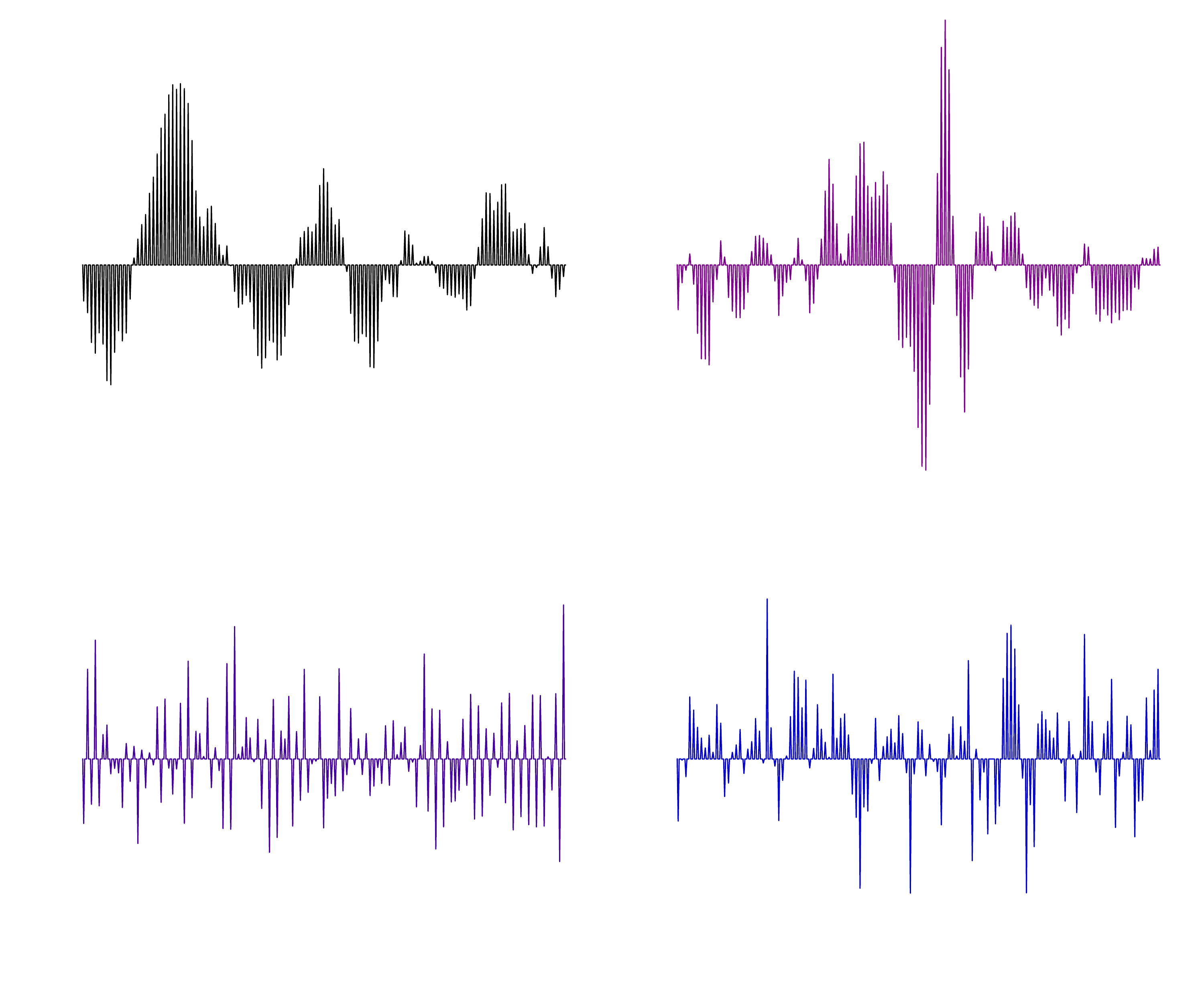}\label{fig:spring-sparse}}\hspace{0ex}\\
\subfloat[]{\includegraphics[width=.49\linewidth, height=0.245\linewidth]{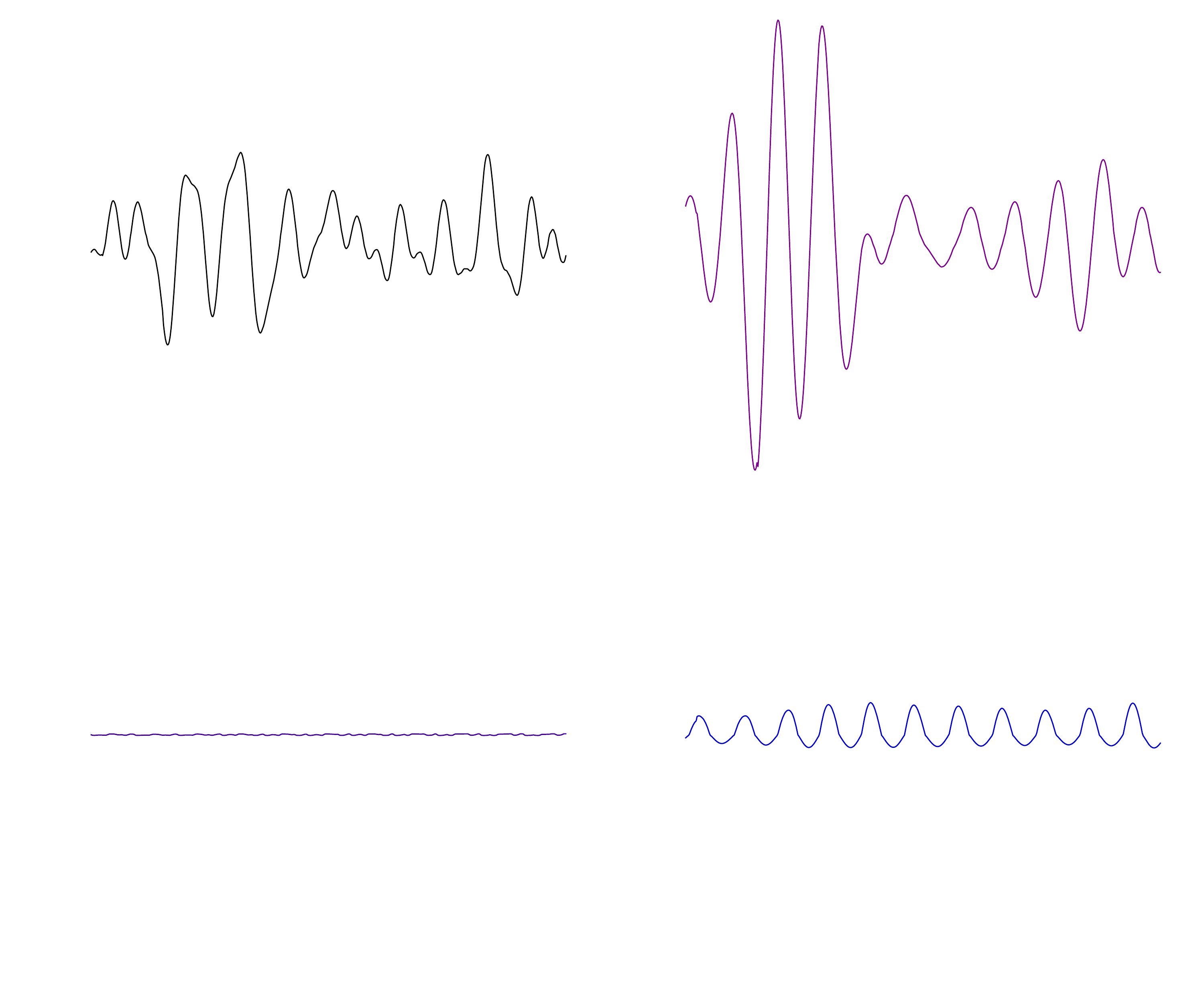}\label{fig:spring-x1.1}}\hspace{0ex}
\subfloat[]{\includegraphics[width=.49\linewidth, height=0.245\linewidth]{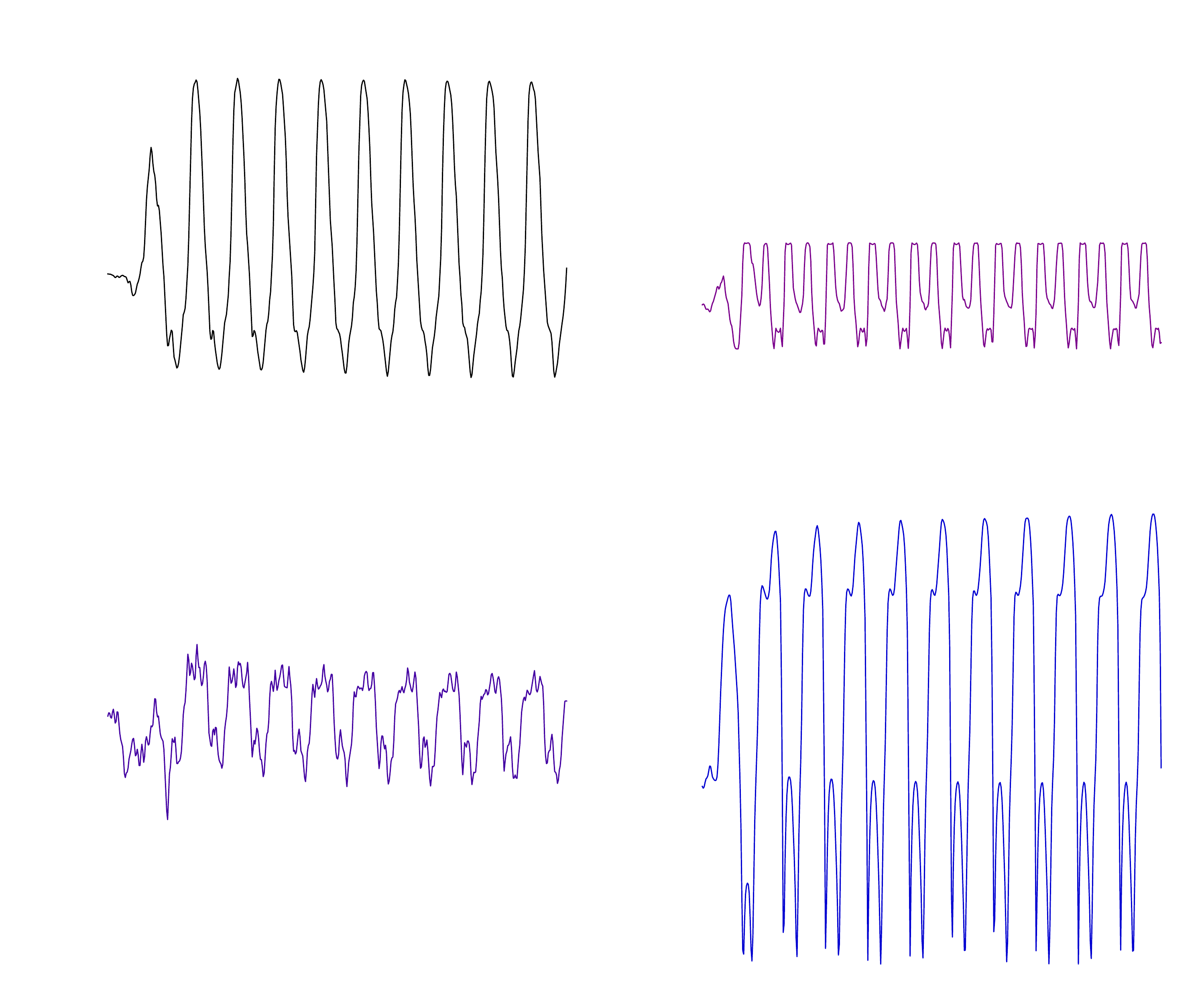}\label{ffig:spring-x1.2}}\hspace{0ex}
\end{center}
\end{minipage}
\end{center}
\caption{Various internal plots for \textit{model-1} from the test dataset of the \textit{spring} modeling task. \ref{fig:spring-x1}) $4$ rows from the frequency band decomposition $\boldsymbol{R}$. \ref{fig:spring-sparse}) From $\boldsymbol{\hat{Z}_{3}}$, corresponding $4$ sparse FIR filters learned by the latent-space. Following the filter bank architecture, \ref{fig:spring-x1.1}) and \ref{ffig:spring-x1.2}) show the corresponding $4$ rows from $\boldsymbol{\hat{X}_{1.1}}$ and $\boldsymbol{\hat{X}_{1.2}}$ respectively. Vertical axes are unitless and horizontal axes are time.}
\label{fig:spring-plots}
\end{figure}

\vspace{-1ex}
In order to compare the performance of the proposed architecture (\textit{model-1}), we use the network from \cite{martinez2019general} (\textit{model-2}), which has proven capable of modeling electromechanical devices such as the Leslie speaker. The latter presents an architecture similar to \textit{model-1}, although its latent-space and back-end have been designed to explicitly learn and apply a modulation in order to match modulation based audio effects \cite{reiss2014audio}. Both models are trained under the same procedure, tested with samples from the test dataset and the audio results are available online\footref{website}. Table \ref{table:loss} shows the corresponding \textit{loss} values. The number of parameters for \textit{model-1} and \textit{model-2} are $410,977$ and $275,073$ and the time each model takes to process a $2$ second audio sample is $0.752$ and $0.4066$ seconds, respectively. This using a \textit{Titan XP GPU} and non real-time \textit{python} implementation.

The proposed model outperforms \textit{model-2} in both tasks. For both reverb tasks and from the test subset, Fig. \ref{fig:frames} shows selected input, target, and output waveforms together with their respective spectrograms.
It can be seen that \textit{model-1} matches very closely the target in the time and frequency domains. From the spectrograms, the smooth noise-like response of the \textit{plate} and the dispersive reflections of the \textit{spring} are noticeable. Overall, the initial onset responses are being modeled more accurately, whereas the late reflections differ more prominently in the case of the \textit{spring}, which across models presents a higher loss. These differences in the frequency domain also correspond to the larger differences between the \textit{mse} and the \textit{mae}, thus, further exploration of the loss weights can be conducted.

Fig. \ref{fig:spring-plots} depicts internal plots of \textit{model-1} when processing the frames from Fig. \ref{fig:frames}c. It shows how the model processes the input frame into the frequency band decomposition $\boldsymbol{R}$ and learns a set of sparse FIR filters $\boldsymbol{\hat{Z}_{3}}$ for each frequency band. Then, the frequency dependent reverberation response $\boldsymbol{\hat{X}_{1.1}}$ is obtained by applying the filters and envelopes to $\boldsymbol{R}$. The nonlinear transformation of the direct sound $\boldsymbol{\hat{X}_{1.2}}$ is accomplished through the learnt waveshapers. These two representations are added together via a time-varying mixing gain, which is fed to the last layer so the audio waveform is reconstructed in the same manner as the front-end that decomposed it.

\vspace{-2.5ex}
\subsection{Listening test}
\vspace{-1ex}

 Thirty participants between the ages of $23$ and $46$ took part in the experiment which was conducted at a professional listening room at Queen Mary University of London. The subjects were among musicians, sound engineers or experienced in critical listening. The audio was played via \textit{Beyerdynamic DT-770 PRO} studio headphones and the Web Audio Evaluation Tool \cite{waet2015} was used to set up the test. 
 
 The participants were presented with samples from the test subset. Each page contained a reference sound, i.e. from the original \textit{plate} or \textit{spring} reverb. Participants were asked to rate $4$ different samples according to the similarity of these in relation to the reference sound. The aim of the test was to identify which sound is closer to the reference. The samples consisted of outputs from \textit{model-1}, \textit{model-2}, a hidden copy of the reference and a dry sample as hidden anchor. Thus, the test is based on the MUSHRA method \cite{international2003ITU}.
 
\begin{figure}[t]
\begin{center}
\begin{minipage}{0.485\textwidth}
\begin{center}
\includegraphics[width=0.75\linewidth]{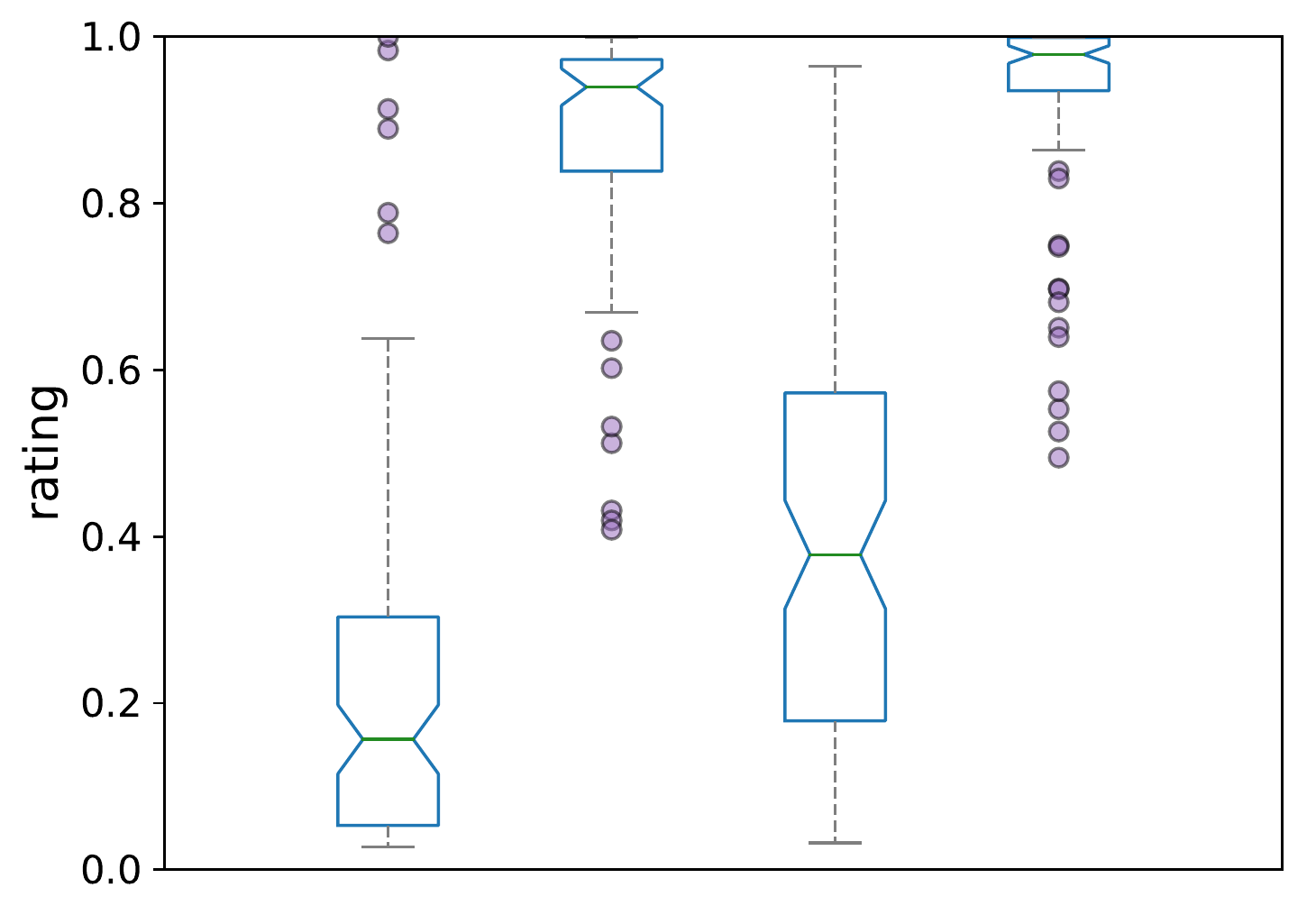}\\
\includegraphics[width=0.75\linewidth]{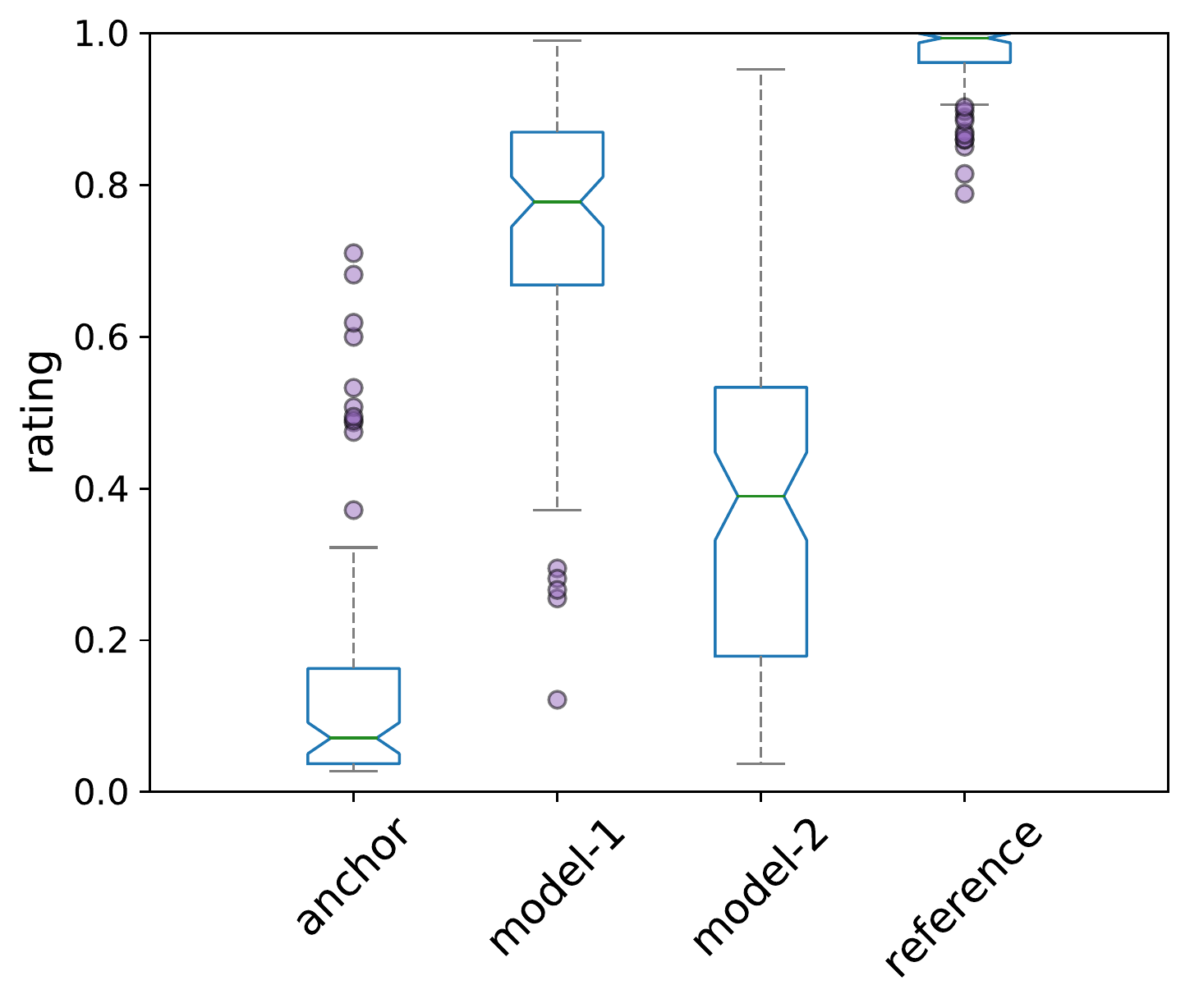}
\caption{\label{fig:boxplot}{Box plot showing the rating results of the listening tests. From top to bottom: \textit{plate} and \textit{spring} reverb tasks.}}
\end{center}
\end{minipage}
\end{center}
\end{figure}

 The results of the listening test can be seen in Fig. \ref{fig:boxplot} as a notched box plot. The end of the boxes represents the first and third quartiles, the end of the notches represents a $95\%$ confidence interval, the green line depicts the median rating and the circles represent outliers. As expected, both anchor and reference have the lowest and highest median respectively. It can be seen that for both \textit{plate} and \textit{spring} reverb tasks, \textit{model-1} is rated highly whereas \textit{model-2} fails to accomplish the reverberation task. Thus, the perceptual findings confirm the results obtained with the \textit{loss} metric and likewise, \textit{plate} models have a better matching that \textit{spring} reverberators. The rating and \textit{loss} values for \textit{spring} do not represent a significant decrease of performance, nevertheless, the modeling of \textit{spring} late reflections could be further explored via a larger number of filters, different loss weights or input frame sizes. 
 
\vspace{-2ex}
\section{Conclusion}
\vspace{-1ex}
\label{sec:conclusion}

In this work, we introduced a signal processing-informed deep learning architecture for modeling artificial reverberators. We explored the capabilities of learning sparse FIR filters and time-varying mixing gains within a DNN framework. We show the model successfully matching nonlinear time-varying transformations such as \textit{plate} and \textit{spring} reverb. Listening test results show that the model emulates closely the electromechanical devices and outperforms other DNNs for black-box modeling of audio effects. As future work, a parametric model, longer decay times and late reflections, and real-time implementations could be explored together with applications beyond effects modeling such as automatic reverberation and mixing.

\bibliographystyle{IEEEbib}
\bibliography{refs}

\end{document}